\documentstyle[12pt]{article}
\input{tcilatex}

\begin{document}

\title{Statistical and Dynamical Study of Disease Propagation in a Small World
Network}
\author{Nouredine Zekri$^{1,2 \thanks{%
Email zekri@mail.univ-usto.dz}}$ and Jean Pierre Clerc$^{1 \thanks{
Email clair@iusti.univ-mrs.fr}}$ \\
%EndAName
{\small $^{1}$I.U.S.T.I., Technop\^{o}le Ch\^{a}teau Gombert, Universit\'{e}
de Provence, Marseille, France. } \\
{\small $^{2}$U.S.T.O., D\'{e}partement de Physique, L.E.P.M., B.P.1505 El
M'Naouar, Oran, Alg\'{e}rie. }}
\maketitle

\begin{abstract}
\hspace{0.33in} We study numerically statistical properties and dynamical
disease propagation using a percolation model on a one dimensional small
world network. The parameters chosen correspond to a realistic network of
school age children. We found that percolation threshold decreases as a
power law as the short cut fluctuations increase. We found also the number
of infected sites grows exponentially with time and its rate depends
logarithmically on the density of susceptibles. This behavior provides an
interesting way to estimate the serology for a given population from the
measurement of the disease growing rate during an epidemic phase. We have
also examined the case in which the infection probability of nearest
neighbors is different from that of short cuts. We found a double diffusion
behavior with a slower diffusion between the characteristic times.
\end{abstract}

\noindent {\bf Keywords:} Small world, percolation, epidemics.

\noindent PACS Number(s): 87.23.Ge, 05.40.-a,05.70.Jk, 64.60.Fr

\newpage

\section{Introduction}

\hspace{0.33in} To model disease propagation, it is necessary to define the
corresponding social network connecting any two individuals in the world.
The expected properties of such a network should be both the clustering
(which excludes models of disorder like the random graphs \cite{Bolob}), and
to allow a connection between any two individuals within a finite number of
steps (which excludes the regular networks with only nearest neighbor
connections). Indeed, for the latter feature Milgram showed in $1967$ that
the average number of steps connecting any two individuals is six (called
also six degrees of separation) \cite{Milgr}. This behavior led recently
Watts and Strogatz to propose the model of Small World Network ($SWN$) \cite
{Watts,Watt2}. They considered a low dimensional network with periodic
boundary conditions for convenience (a ring for example) where they rewired
some bonds with a probability $\phi $ to a new site randomly chosen from the
network. For small values of $\phi $ this still corresponds to a regular
network but with few long range connections called Short-Cuts ($SC$). A more
recent work on the $SWN$ was proposed by Newman and Watts \cite{Newma} where
the number $k$\ of Nearest Neighbors ($NN$) is conserved but instead of
rewiring, they added an average density $\phi $ of new bonds from each site $%
i$ to other nodes randomly chosen (except its nearest neighbors). A review
of these models and their application to various fields and particularly
epidemics can be found in the references \cite{Watts,Newm2}. In these
networks the percolation threshold was extensively investigated and its
dependence to the $NN$ and $SC$ was found to satisfy the following equation 
\cite{Newma,Newm3} 
\begin{equation}
\phi =\frac{(1-p_{c})^{k/2}}{p_{c}}
\end{equation}

\noindent This threshold corresponds in epidemics to the smallest
concentration of susceptibles leading to the outbreak \cite{Newma}. However,
the statistical behavior of percolative $SWN$ networks is different from
that of regular systems \cite{Stauf}. In particular, at the percolation
threshold, there is no diverging cluster for the $SWN$ because a $SC$
between the two ends of the system has a finite probability to occur for any
non-vanishing number of these bonds. On the other hand, the characteristic
length scale in such networks (which corresponds to the correlation length
in regular lattices) behaves as $\phi ^{-1/d}$, $d$ being the euclidean
dimension of the system \cite{Newm3}. It is then obvious that this
characteristic length does not diverge at the percolation threshold for such
networks. Therefore, a further investigation of the cluster statistics
around this {\it new} percolation threshold for such networks and the phase
transition seems to be necessary.

\hspace{0.33in} Let us now consider the application of this model to
epidemics which seems to be one of its main aims. From the large amount of
works using $SWN$, there is no direct comparison with the existing data may
be due to the complexity of the diseases features (incubation and latent
periods, birth and death rates etc.). Furthermore, the parameters used
(mainly $\phi $) are very small and do not simulate the real connections
between individuals. They use also commonly average values of the $NN$ and
the $SC$ while these quantities strongly fluctuate in the real live (the
number of contacts, friends, family members etc. varies from $0$ to few
tens) which can influence sensitively the results on the density of
susceptibles at the percolation threshold (epidemic outbreak). On the other
hand, it is impossible in practice to measure the density of susceptibles
systematically (it needs an extensive serological investigation in the
epidemic phase). Generally, for large population samples epidemiologists
measure the evolution with time of the number of cases for a given disease.
It is then necessary to study the dynamical behavior of the propagation of
the disease and relate it to the density of susceptibles. There are only few
works which examined (only qualitatively) the dynamical behavior of the
disease on social networks \cite{Pasto, Moore, Abram}. The aim of these
works was to show how the density of infected behaves in the endemic and
epidemic phases.

\hspace{0.33in} In this article, we use a site percolation on a $SWN$ with
parameters ($k$ and $\phi $) representing a sample of school age children to
study the effect of the fluctuations of $NN$ and $SC$ on the percolation
threshold. Furthermore, in order to propose a formula for determining the
serology of the sample from the rate of increase of the number of cases, we
investigate also extensively the dynamical behavior of an infectious disease
as well as its effect on the density of susceptibles below and above the
percolation threshold. A new {\it super-diffusion} is found above the
percolation threshold when the cluster is initially infected by one or a
small number of infectious sites and its characteristic time dependence on
the density of susceptibles is determined. We examined also the case where
the infection probability of the $NN$ is different from that of $SC$,
showing a double diffusion with two characteristic times. In the next
section we describe the model and then present the results on the statistics
of the clusters and the percolation threshold. The results on the dynamical
behavior of the disease are presented in section 4.

\section{Model description}

\hspace{0.33in}We consider the one dimensional $SWN$ described by Newman and
Watts \cite{Newma} but $\phi $ represents the total number of $SC$ generated
for each site uniformly from all the other sites of the network. In the case
where $k$ and $\phi $ are not fixed they are generated randomly within a
normal distribution centered at their average values with fluctuations $%
\delta k$ and $\delta \phi $ respectively. The coordination number is the
total number of bonds to a given site ($z=k+\phi $). We study in this
network a site percolation problem \cite{Stauf} by assuming each susceptible
site $j$ (occupied) contracts the disease if it is connected with an ill
site $i$ (occupied also). The occupied sites (susceptibles) are randomly
generated with a concentration $p$ while the empty sites correspond to
refracted individuals. For $k$ and $\phi $ fixed, the percolation threshold $%
p_{c}$ is related to $k $ and $\phi $ by Eq.(1) \cite{Newm3}. This threshold
corresponds to a transition from the endemic phase below $p_{c}$ to the
epidemic one above this point \cite{Moore}. In $SWN$ networks, $p_{c}$ is
the minimum concentration of occupied sites above which the average largest
cluster size $\xi $ of the occupied sites becomes power-law increasing with
the concentration ($\xi $ $=(p-p_{c})^{x}$ ), while it diverges in a regular
network \cite{Stauf} (note here that the exponent $x$ is positive). By
analogy with the regular lattices \cite{Stauf}, we will check the
universality of the exponent $x$.

\hspace{0.33in}We are interested in the application of such a model to a
childhood disease like measles. In such diseases epidemiological
investigations on school age children can be easily controlled and provide
data with a minimum bias. We choose parameter values corresponding to such a
disease by taking $k=2$ to be the average number of brothers, sisters and
neighbors, while $\phi =30$ represents the average number of children one
can meet at the school. These parameters should correspond to a topology
closer to that encountered in a real social network. Regarding the dynamical
study, we assume the major contribution to the epidemics provided by the
largest cluster. We restrict ourselves then to this cluster and start the
infection with one or few infectious sites at time $0$. These sites will
infect all the connected sites in the next step (after a time $\Delta t$),
which themselves infect their connected sites after $2\Delta t$ and so on.
We assume the latent and incubation periods smaller than $\Delta t$ which is
taken in the rest of this paper as a unit time. The number of infected sites
in each step is averaged by varying the initial infectious site position
through the whole cluster.

\section{Percolation threshold and cluster distribution}

\hspace{0.33in} In this section, we realize $100$ configurations of the
network described in the previous section with a size fixed at $100000$
sites. We examine the effects of $\phi $ and its fluctuations on the average
cluster sizes, $p_{c}$ and $x$. Finally, we investigate the cluster size
distribution around $p_{c}$ in order to determine the main contribution to
the propagation of the disease.

\hspace{0.33in} In figure 1a we show the variation of the cluster size with
the concentration of occupied sites for three different cases: $\phi =6,30$
(fixed values) and for $k$ and $\phi $ randomly generated with a normal
distribution centered at $2$ and $30$ respectively, with a fluctuation of $2$
et $15$ respectively. We see clearly from this figure that in all cases the
cluster sizes vary as a power law of ($p-p_{c}$ ) above $p_{c}$. For fixed $%
k $ and $\phi $ the value of $\ p_{c}$ is in a good agreement with the
analytical predictions of Newman et al. \cite{Newm3} (Eq. 1). However, in
the case of fluctuations of $k$ and $\phi $, this threshold decreases
sensitively (about $50\%$ in the case shown in Fig.1a) as the fluctuations
increase. Therefore, the average values of $k$ and $\phi $ are not
sufficient to characterize an epidemic outbreak. The $SC$ fluctuations $%
\delta \phi $ decrease $p_{c}$ as a power law with an exponent $0.1$ (as
shown in figure 1b), indicating a sensitive participation of the larger
values of $\phi $ to built the largest cluster. Therefore, the percolation
threshold behaves as 
\begin{equation}
p_{c}\simeq \phi ^{-1}\delta \phi ^{-0.1}
\end{equation}

\noindent From this behavior, we can estimate the percolation threshold in a
real sample of school age children to be in the range $2.3\%$ to $2.8\%$.

\hspace{0.33in} Now let us restrict ourselves to the case of fixed $k=2$ and 
$\phi =6$ in order to examine the statistical behavior of the clusters
around $p_{c}$ (without loss of generality, these values are chosen only
because $p_{c}$ is large enough to enable sufficient cluster statistics for
such a sample size) . We found that the cluster size fluctuations are
maximum at this threshold (see Fig.2a) implying a divergence of this
quantity at $p_{c}$ which seems to be a signature of a phase transition. The
cluster size distribution (see figure 2b) confirms this divergence since it
decreases exponentially below $p_{c}$ while it is power-law decreasing at
this threshold (this power law behavior is in agreement with the results of
Castellano et al. \cite{Caste} on other systems). Indeed, at $p_{c}$ this
corresponds to a L\'{e}vy distribution \cite{Levy} with an exponent of $2.13$
indicating the divergence of all its moments. We notice here that only the
higher sizes (rare events) contribute to the outbreak at $p_{c}$ (as
expected in such distributions). Above $p_{c}$ the small size clusters are 
{\it absorbed} by the largest one and we have again an exponentially
decreasing distribution for small clusters while there is only one very
large cluster (not shown in Fig.2b).

\hspace{0.33in} Since the cluster size does not diverge at $p_{c}$, it is
obvious that $x$ is not universal (because it is not a critical exponent),
but it is interesting to know how it depends on $\phi $ in such lattices. In
figure 2c, the exponent $x$ seems to vary only linearly for larger values of 
$\phi $ but with a very small slope (about $5.6$ $10^{-3}$). It is difficult
to predict its behavior for very small values of $\phi $ because in this
case the network tends to a regular one and the cluster size becomes very
large so that the sample sizes taken here do not allow us to measure this
exponent accurately.

\hspace{0.33in} However, even if the parameters chosen in this model are
close to those of a real social network, it seems impossible for
epidemiologists to check these results. Indeed, as explained below, they
cannot measure the density of susceptibles, except if they investigate
systematically the serology of a sufficiently large sample of school age
children (e.,e.g. for a city sample). Therefore, the behavior of $p_{c}$
should be checked for measurable quantities. In the case of disease
propagation, the time dependence of the number of cases can be directly
measured by epidemiological techniques. We will investigate this dynamical
behavior in the following section.

\section{Dynamical study of the propagation of a disease}

\hspace{0.33in} In this section we restrict ourselves to the fixed values of 
$k$ and $\phi $ ($2$ and $30$ respectively) to simulate a sample of school
age children. From the results of Fig.2b, we assume that the main growing
effect of the infection comes from the largest cluster and estimate the
propagation time of the epidemics only from this cluster. We determine the
evolution with time of the number of cases for both phases endemic ( $%
p<p_{c} $ ) and epidemic ( $p\geq p_{c}$ ). As found in figure 2a the
cluster size at $p_{c}$ strongly fluctuates and therefore, the time behavior
of the number of cases also fluctuates. The variation of the number of cases
with time is shown in figure 3a for three cases (just below $p_{c}$ , at $%
p_{c}$ and above $p_{c}$) with only one initial infectious site. In both
cases, the number of cases increases up to a maximum and then decrease
because the number of susceptibles decreases. In the endemic phase, the
number of connections between occupied sites in the cluster is mostly $1$
and does not allow a significant increase of the number of cases (the
behavior in this case is under estimated since all the clusters should
contribute to this increase). For susceptible densities around $p_{c}$ this
situation persists for a long time and the number of cases increases
linearly with time showing a normal diffusion of the disease. In the
epidemic phase the increase becomes exponential indicating a new kind of 
{\it super-diffusion} \cite{Levy, Evang} of the disease, due mainly to the
increasing number of connections in the cluster (as shown in figure 3b).
This exponential growth is also observed for $SIR$ models \cite{Ander} where
the rate is proportional to the basic reproduction rate $R_{0}$ which
correponds in our case to the average number of connections in the cluster.
We have also performed a Monte-Carlo simulation to the measles propagation
in a more realistic sample (births, deaths, incubation and latent periods
etc.) where the average infections is $2$ for each infectious individual and
found also an exponential growth of the infected cases \cite{Zekri}.
Therefore, this exponential growth does not seem to depend on the topology
of the sample but the rate is sensitive to the geometry of the network. Note
in the present work that in the case of more than one initial infectious site
(see figure 3c) the exponential growth behavior does not change but the
growing rate fluctuates due to the fluctuating number of connections. The
average rate of the exponential growth $\gamma $ (corresponding to the
characteristic time of the epidemics) is shown in figure 4 to increases as $%
Ln(p)$ above $p_{c}$ while the period of this epidemic behavior decreases.
From this figure we can conclude that when the characteristic time decreases
below $5$ (or $\gamma $ increases above $0.2$), the epidemic behavior takes
place. This behavior seems to have a direct application in epidemiology
since it provides a method for the estimation of the serological situation
(density of susceptibles) from the characteristic time which is easily
measurable. Therefore, this result stimulates a proposal for a serological
examination for a given childhood disease in a sample of age school
children, but during an epidemic period to compare a realistic behavior with
that obtained in this paper.

\hspace{0.33in} Now let us examine the case of adding different infection
probabilities to this system. We consider that a site $i$ infects another
site $j$ with a probability $p_{n}$ if $j$ is a neighbor of $i$ and $p_{sc}$
if it is a short cut. The motivation of this investigation is that a
susceptible child has a different probability to be infected by his brothers
(or sisters) than by the other children meeting him at the school. We see
clearly a double diffusion behavior in figure 5 (for $p_{n}=0.1$ and $%
p_{sc}=0.9$), where the number of infected starts growing exponentially up
to the characteristic time ($1/\gamma $), then it increases as a power law
up to a new characteristic time from which it grows again exponentially with
the same rate. The slow diffusion is due to the small contact probability
for the neighbors ($p_{n}=0.1$) and has been observed in other fields \cite
{Dykhn}. This slow diffusion appears very short because the number of $NN$
is very small ($k=2$). It should be interesting to investigate this double
diffusion for larger $k$ (which is the case of animal diseases).

\section{Conclusion}

\hspace{0.33in} We have investigated in this article the statistics of the
cluster sizes in a one dimensional $SWN$ by taking into account the $NN$ and 
$SC$ fluctuations. We found that these fluctuations decrease $p_{c}$ as a
power law with a small exponent leading to a new expression for the
percolation threshold. We found also that cluster size fluctuations is the
quantity governing the phase transition in such a network. On the other
hand, in order to apply our results to the measured quantities in
epidemiology, we have studied the dynamics of the disease propagation in
such clusters. We found in epidemic phases a {\it super-diffusive} with an
exponentially growing number of infected sites, while at $p_{c}$ this number
increases as a power law. The growing characteristic time is larger than $5$
in the endemic situation and decreases linearly with $Ln(p)$ in the epidemic
phase. This result provides a way to estimate the density of susceptibles in
the epidemic phase. We propose then a serological investigation in epidemic
situations to check this behavior. Finally, we examined the case where the
infection probability is very small in the $NN$ compared to the $SC$. The
dynamical behavior of infected cases shows a double diffusion with two
characteristic times, and a power-law increase (deceleration) between them.
We think that this effect is useful for samples with large $NN$ and shows a
way to stop the propagation of the epidemic for other diseases.

\vspace{0.2 in}

{\bf ACKNOWLEDGEMENTS}

\hspace{0.33in} One of the authors (NZ) would like to thank the Arab Fund
for Economic and Social Development for the financial support and the staff
of the I.U.S.T.I., Universit\'{e} de Provence - Marseille for hospitality
during the progress of this work. We thank Professor M.Barth\'{e}l\'{e}my
for fruitful discussions and Professor A.M.Dykhne for drawing our attention
to the double diffusion behavior.

\newpage

{\Large {\bf Figure Captions}}

\bigskip

{\bf Figure 1\qquad a) } Cluster size (number of sites in the cluster) versus concentration of the occupied
sites for three cases: $\phi =6$ (solid curve); $\phi =30$ (dotted curve)
and $\phi =30$ with fluctuations $\delta k=2$, $\delta \phi =15$
(dahs-dotted curve).

\bigskip \qquad \qquad\ \ \ \ \ \ \ {\bf b) } $p_{c}$ versus $\delta \phi $ (sites) 
for $\phi =30$ sites. The solid line is a fit of the data.

\bigskip {\bf Figure 2\qquad a)} Cluster size fluctuations (sites) versus the
occupied sites concentration ($\phi =6$ sites). The solid curve is a guide for the
eyes. 

\bigskip \qquad \qquad \qquad {\bf b)} Distribution of the cluster size (sites) for
fixed ($k=2$, $\phi =6$) in a semi-log plot \newline
at $p_{c}$ (solid curve) and $p=1\%$ (dotted curve). The dotted line is a
linear fit of the data below $p_{c}$. The insert is a log-log plot of the
distribution at $p_{c}$ with a linear fit. \bigskip

\qquad \qquad \qquad {\bf c}) Variation of the exponent $x$ with the number
of short cuts $\phi $ (sites). The solid line is a linear fit of the data to $5.6$ $%
10^{-3}$.

\bigskip

{\bf Figure 3\qquad a)} Number of cases versus time for three different
cases; $p=3.5\%$ (solid curve), $p=4.5\%$ (dashed curve) and $p=8\%$ (dotted
curve). Insert: log-log plot with a power law fit of $p=4.5\%$ and an
exponential fit of $p=8\%$.

\bigskip \qquad \qquad \qquad {\bf b)} Distribution of the number of
connections (acquaintances) in the largest cluster for $p=3.5\%$ (solid curve), $p=5\%$
(dashed curve) and $p=10\%$ (dotted curve).

\bigskip \qquad \qquad \qquad {\bf c)} The rate of the exponential growth
(in arbitrary units) versus number of initial infectious sites. The horizontal line is the
average rate.

\bigskip

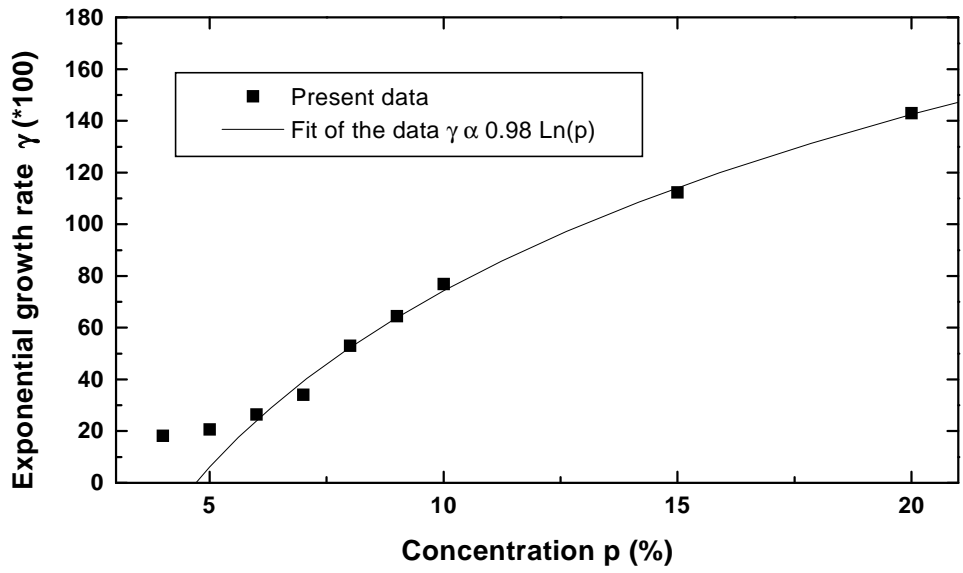
{\bf Figure 4} \qquad The rate of the exponential growth versus $p$. The
solid line is a fit of the curve linearly with $Ln(p)$.

\bigskip

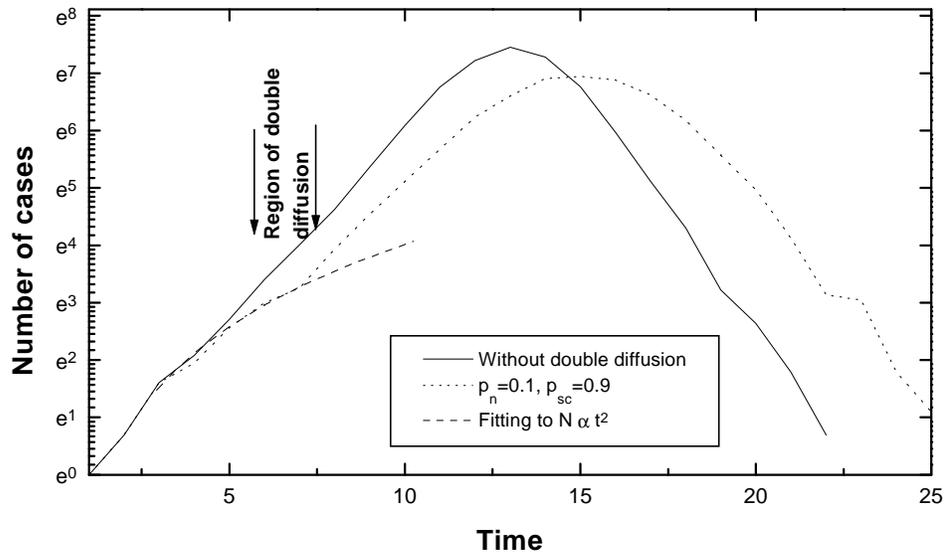
{\bf Figure 5} \qquad The Number of infected cases versus time (in arbitrary units) for one
initial infectious site and an infection probability one (solid curve), and the
probabilities of infection: $p_{n}=0.1$ and $p_{sc}=0.9$ (dotted curve). The
dashed curve is a power law fit of the second data in the region of the
double diffusion.

\bigskip

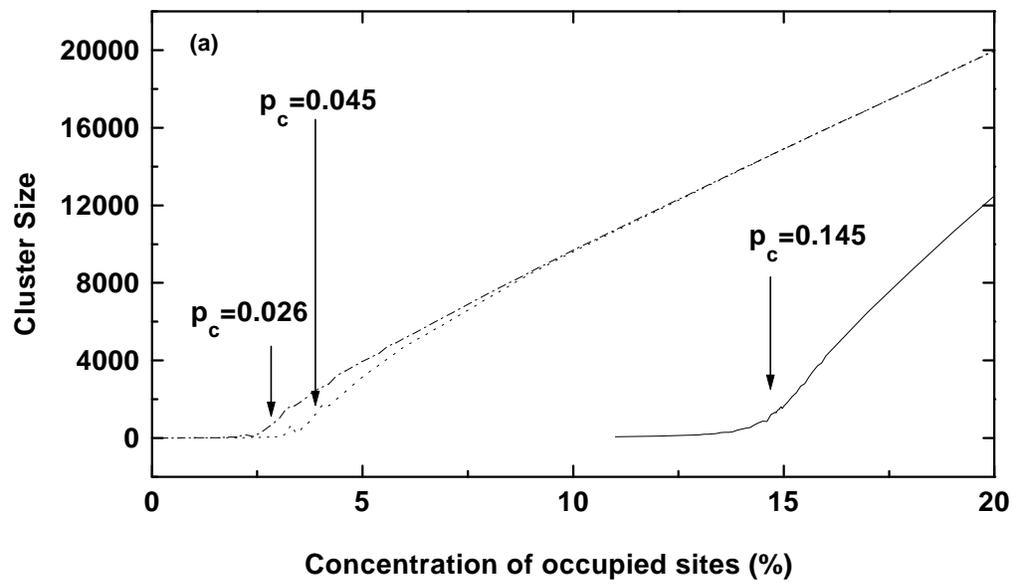
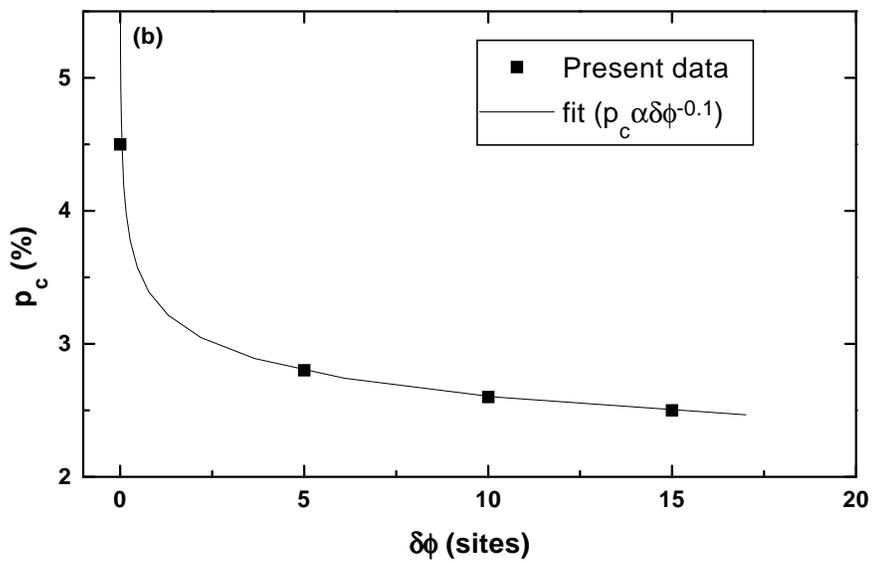
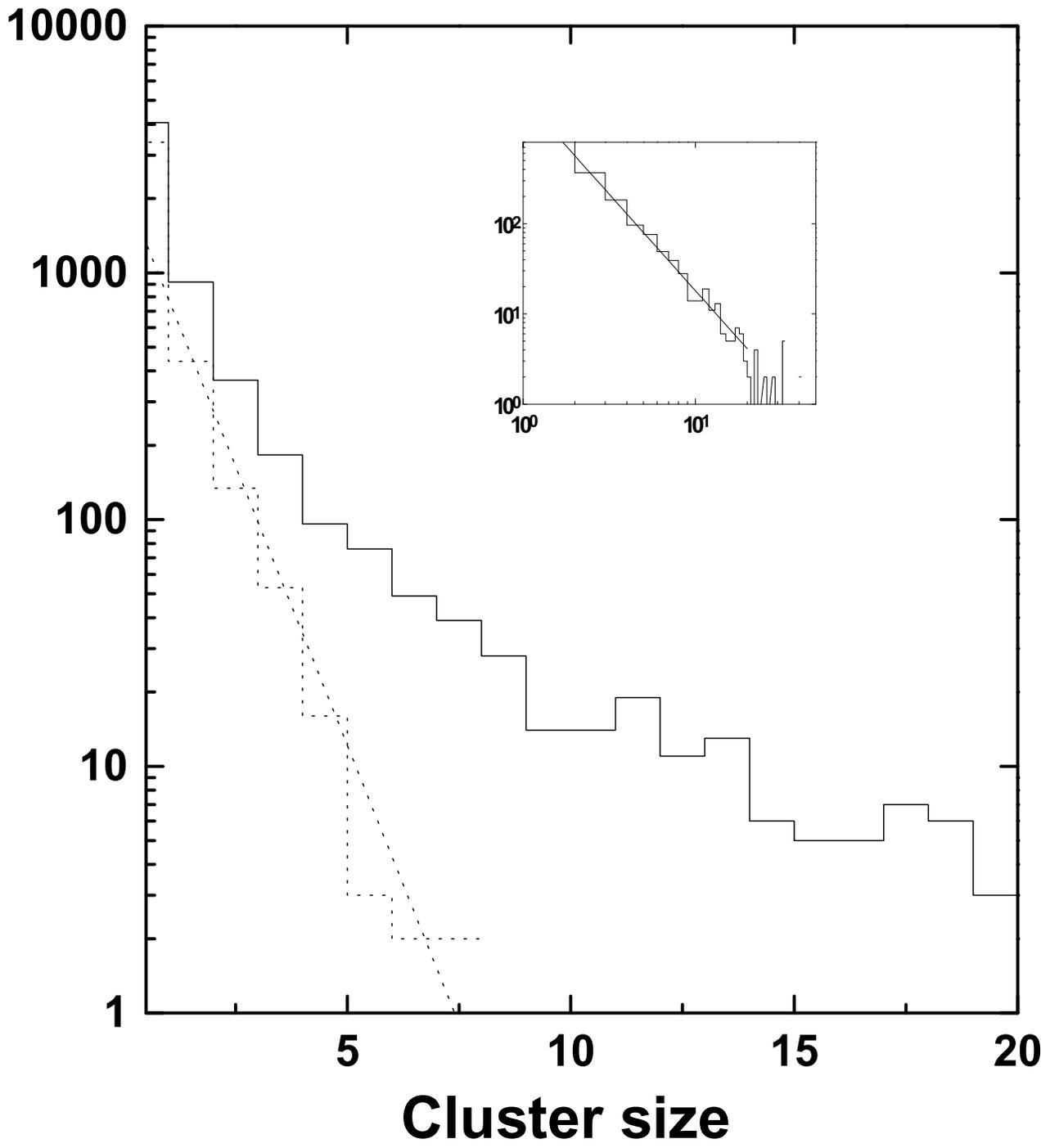
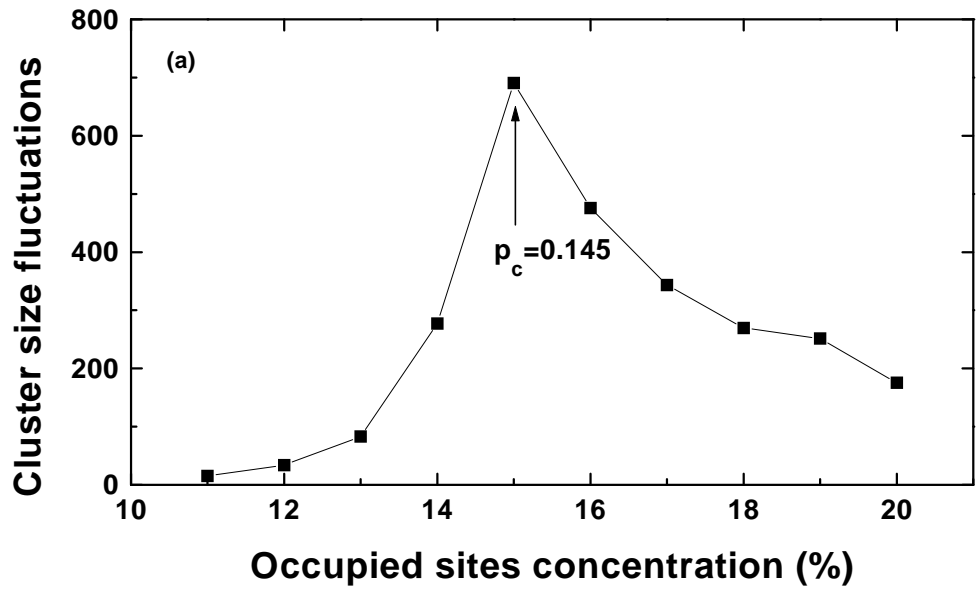
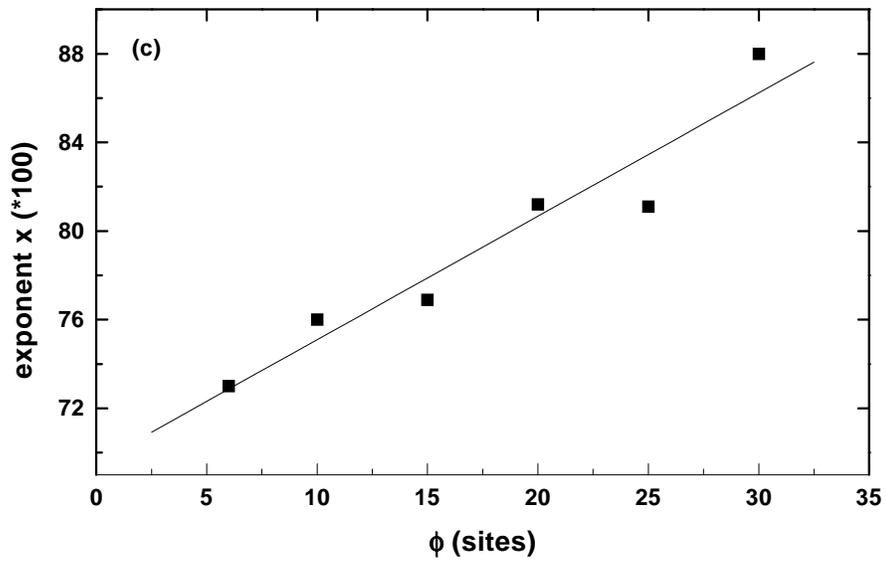
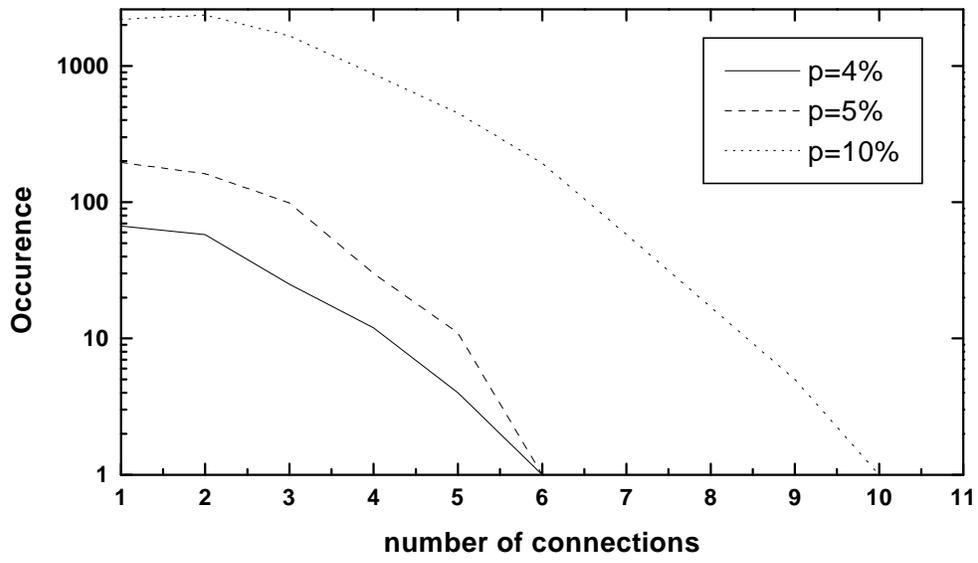
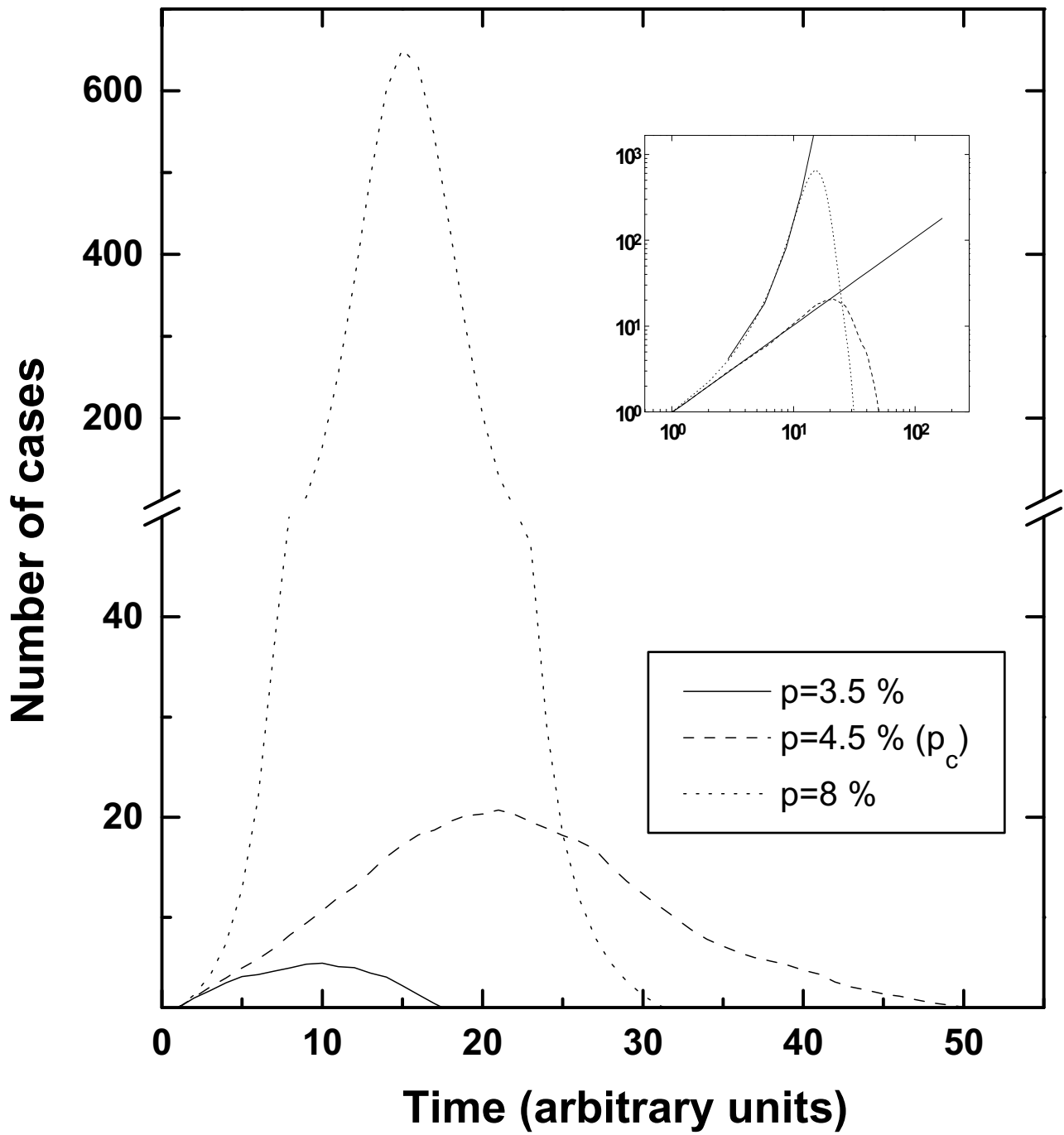
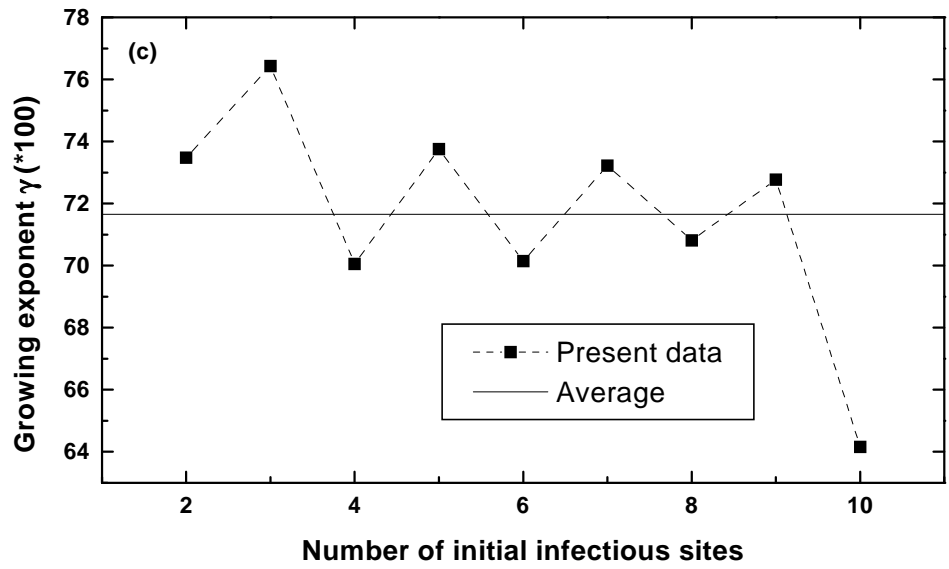
\end{document}